\documentclass[twocolumn,showpacs,superscriptaddress,preprintnumbers,amsmath,amssymb,prl]{revtex4}

\usepackage{graphicx}	
\usepackage{dcolumn}	
\usepackage{bm}	
\usepackage{color}	
\usepackage{multirow}	
\usepackage{epsfig}

\usepackage{amsmath}
\usepackage{amssymb}

\newcommand     {\beq}[1]         { \begin{equation} #1 \end{equation} }

\begin{document}

\title{New universality class for the fragmentation of plastic
materials}

\author {G.\ Tim\'ar}
\affiliation{Department of Theoretical Physics, University of
  Debrecen, P. O. Box:5, H-4010 Debrecen, Hungary}
\affiliation{Computational Physics IfB, HIF, ETH,
  H\"onggerberg, 8093 Z\"urich, Switzerland}
\author{J.\ Bl\"omer}
\affiliation{Spezialwerkstoffe, Fraunhofer UMSICHT
Osterfelder Str.\ 3, 46047 Oberhausen, Germany}
\author {F.\ Kun}
\affiliation{Department of Theoretical Physics, University of
  Debrecen, P. O. Box:5, H-4010 Debrecen, Hungary}
\author {H.\ J.\ Herrmann}
\affiliation{Computational Physics IfB, HIF, ETH,
  H\"onggerberg, 8093 Z\"urich, Switzerland}
\affiliation{Departamento de Fisica, Universidade Federal do Ceara,
  60451-970 Fortaleza, Ceara, Brazil}
\date{\today}	

\begin{abstract}
We present an experimental and theoretical study of the fragmentation
of polymeric materials by impacting polypropylene particles of spherical
shape against a hard wall. Experiments reveal a power law mass
distribution of fragments with an exponent close to $1.2$, which is
significantly different from the known exponents of three-dimensional
bulk materials. A 3D discrete element model is introduced which
reproduces both the large permanent deformation of the polymer during
impact, and the novel value of the mass distribution exponent. We
demonstrate that the dominance of shear in the crack formation and the
plastic response of the material are the key features which give
rise to the emergence of the novel universality class of fragmentation
phenomena. 
\end{abstract}


\maketitle

Fragmentation phenomena are ubiquitous in nature and play a crucial
role in numerous industrial processes related to mining and ore
processing \cite{astrom_advinphys_2006}. 
A large variety of measurements starting from the
breakup of heavy nuclei through the usage of explosives in mining or
fragmenting asteroids revealed the existence of a
striking universality in fragmentation phenomena
\cite{astrom_advinphys_2006,turcotte_1986,oddershede_prl1993,meibom_prl1996,astrom2004phys,holian_prl2000,wittel_prl_2004,wittel_shape_prl,katsuragi_prl_2005,glassplate_kadono_1997}:
fragment mass distributions exhibit a power law decay,
independent on the type of energy input (impact, explosion, ...), the
relevant length scales or the dominating microscopic interactions
involved. Detailed laboratory
experiments on the breakup of disordered solids have revealed that
mainly the effective dimensionality of the system determines the value
of the exponent, according to which universality classes of
fragmentation phenomena can be distinguished. Several possible
mechanisms have been put forward to understand the emergence of the
universal power law behavior.
For rapid break-up of heterogeneous bulk solids with a high degree of
brittleness, the self-similar
branching-merging scenario of propagating unstable cracks governed by
tensile stresses  
can explain the main features of the fragment mass 
distribution \cite{astrom_branching_1997,astrom2004phys,astrom_dynfrag_2004,PhysRevE_76_026112_astrom}, while for shell systems an additional sequential binary
breakup mechanism has to be taken into account
\cite{wittel_prl_2004,wittel_shape_prl}.
It is an important question of broad scientific and
technological interest how plasticity,
and the emergence of complicated stress states like shear
affect the fragmentation process. The fundamental questions of how
robust the universality classes are with respect to mechanical
properties and whether there exist further universality classes of
fragmentation of solids, still remain open.

In the present Letter we investigate the
fragmentation process of plastic materials by impacting 
spherical particles made of polypropylene (PP) against a hard wall.
Our experiments show that the mass distribution of plastic fragments
exhibits a power law behavior with an exponent close to 1.2, which is
substantially different from the one of bulk brittle materials in
three-dimensions. In order to
understand the physical origin of the low exponent,
a three-dimensional discrete
element model is developed where the sample is
discretized in terms of spherical particles connected by elastic
beams. To capture the fracture mechanisms of plastic materials in the
model, broken particle contacts are able to reconnect when compressed
against each other leading eventually to the healing of
cracks and to a plastic energy dissipation. By adding two novel
ingredients into a DEM model, namely healing of broken bonds and
breaking under compression, we are able to simulate the plastic
behavior of PP and reproduce in fact the observed new power law
distribution.  
\begin{figure}
\begin{center}
\epsfig{bbllx=20,bblly=20,bburx=575,bbury=250,file=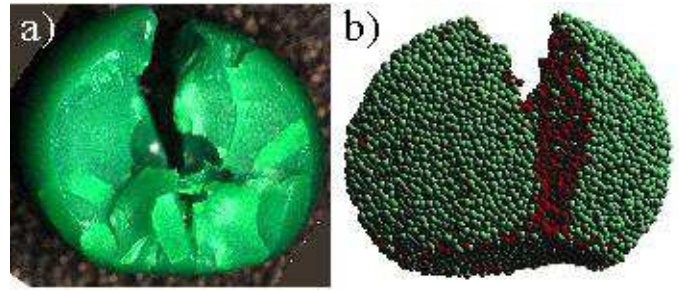,
  width=9.0cm}
 \caption{\small {\it (Color online)}
Final states of impact at low impact velocities in the experiment
$(a)$ and in the simulation $(b)$. In the contact area with the hard
wall large
permanent deformations occur due to compression, while above it
vertical cracks are formed due to tensile stresses. 
The simulations are in very
good agreement with the measurements. 
}
\label{fig:exp_deform}
\end{center}
\vspace*{-0.6cm}
\end{figure}

In the experiments we used isotactic polypropylene, which is a
thermoplastic polymer with an intermediate level of crystallinity in
its molecular structure. The most important
parameters of PP are: Young modulus $1300$ $MPa$ (room
temperature), glass transition temperature $-10$ ${}^oC$, melting
point $160$ ${}^oC$, and density $0.9$ $g/cm^3$. 
Besides tacticity, the mechanical response and fracture
characteristics of PP are also strongly affected both by the
temperature and by the rate of loading: 
increasing strain rate gives rise to a more brittle response while the
raising temperature enhances ductility \cite{plastic_book_super}. 
To achieve fragmentation a single particle
comminution device was used which accelerates particles one-by-one
by centrifugal force in a rotor up to the desired velocity. The
rotor ensures that the particles hit the hard wall at a rectangular
angle in an evacuated environment eliminating the disturbing effect of
inclined impact and of turbulent air flow.
Spherical PP particles of diameter $d=4$ mm were fragmented 
at different impact velocities $v_0$ in the range $30$ m/s-180 m/s.
Figure \ref{fig:exp_deform}$(a)$ shows that at low enough
velocities, the collision does not result in a breakup, 
instead the particles undergo a large plastic deformation at the impact
site. Above the completely flattened contact zone of
permanent deformation meridional cracks form due to tensile stresses,
however, the body does not fall apart, the plastic zone retains its
integrity. 

\begin{figure}
\begin{center}
\epsfig{bbllx=10,bblly=465,bburx=350,bbury=750,file=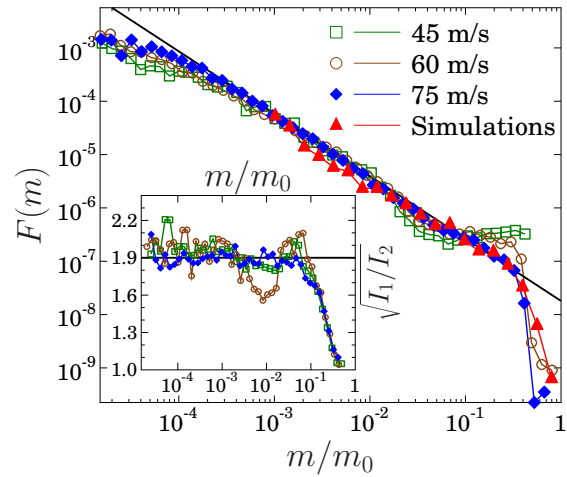,
  width=7.5cm}
 \caption{\small
{\it (Color online)} Mass distribution of fragments obtained from
experiments at three 
different impact velocities. $m_0$ denotes the average mass of PP
spheres. At the highest impact velocity $v_0=75$
m/s, $F(m)$ shows a power law behavior over 3 orders of magnitude followed by
an exponential cutoff for the  very large pieces. Simulation results
obtained with the parameters $\Theta_{th}=1$ and $t_h=0$ ({\it
  bending\&healing}) are in a very 
good agreement with the experimental findings. Inset: the shape
parameter $\sqrt{I_1/I_2}$ of fragments as a function of mass for the
experiments of the main figure.}
\label{fig:exp_mass}
\end{center}
\vspace*{-0.5cm}
\end{figure}

Fragmentation occurs when the impact velocity $v_0$ exceeds a
material dependent critical value $v_c$, which is about $60$m/s for our
PP particles. At each impact velocity
400 particles were fragmented accumulating the fragments 
in the grinding chamber of the machine. In the data analysis,
$99-99.5\%$ of the total mass of the samples was recovered.
In order to evaluate the mass distribution of the fragments, we scanned
the pieces with an open 
scanner obtaining digital images where fragments appear as white spots
on the black background \cite{wittel_prl_2004,wittel_shape_prl}. This way the identification of fragments is
reduced to cluster searching of white pixels. The very fine powder of
extension smaller then the pixel size of the scanner was left out of the
data analysis. The two-dimensional
projected area $w$ of fragments is determined as the number of pixels
of the clusters, from which the mass $m$ of fragments can be estimated
as $m \sim w^{3/2}$ since the three-dimensional fragment shape
is close to isotropic. We did check the shape isotropy by calculating
the square root of the ratio of the larger $I_1$ and smaller $I_2$
eigenvalues of the tensor of inertia of 
the 2D projections. The inset of Fig.\ \ref{fig:exp_mass} shows that
the value of $\sqrt{I_1/I_2}$ is close to 1.9 for almost all fragment
masses indicating a high level of isotropy. The mass distribution
$F(m)$ of fragments 
obtained at three different impact velocities is presented in Fig.\
\ref{fig:exp_mass}. 
It can be observed in the figure that 
at the highest impact velocity $v_0=75$ m/s, where the state of complete
breakup is reached, a power law functional form emerges  
\beq{
F(m) \sim m^{-\tau_{pl}}
}
over more than 3 orders of magnitude in the regime of small
fragments. At lower impact velocities the
power law regime of the distribution is followed by a hump for the
largest fragments which gradually disappears and the cutoff becomes
exponential as $v_0$ increases.
The most astonishing feature of
the experimental results is that the value of the exponent
$\tau_{pl} = 1.2\pm 0.06$ of the power law regime is significantly
lower than the values 
$\tau_{br} \approx 1.8-2.1$ typically found in the fragmentation of
three-dimensional bulk objects consisting of disordered brittle materials
\cite{meibom_prl1996,oddershede_prl1993,holian_prl2000,astrom2004phys,astrom_dynfrag_2004}.
The anomalously low value of $\tau_{pl}$ is the consequence of the
breakup mechanism of plastic materials which has not been considered
by the usual theoretical approaches \cite{astrom_dynfrag_2004,holian_prl2000}.  

In order to reveal the underlying physical mechanisms of the
fragmentation of plastic materials, 
we used a Discrete Element Model (DEM) to simulate the
fragmentation of polymeric particles of spherical shape when they
impact a hard wall. The spherical sample is represented as a random
packing of spheres with a narrow size distribution. 
The particles are connected by beams along the edges of a
Delaunay triangulation of the initial particle positions such that in
the initial state neighboring elements are in contact.
In 3D the total deformation of a beam is
calculated by the superposition of elongation, torsion, as well as
bending and shearing in two different planes \cite{kun_fragment_pre2008}.
To capture one crucial aspect of the deformation behavior and the fracture
of plastic materials, our DEM model has two novel
components, i.e.\ the form of the breaking criterion of the
beam elements and the reactivation of broken contacts under pressure.
The beams break when they get overstressed
according to a physical breaking rule \cite{hjh_prb_fractdis_1989}
\begin{equation}
\frac{\varepsilon |\varepsilon|}{\varepsilon_{th}^2} +
\frac{\text{max}\left( |\Theta_i|,|\Theta_j| \right)
}{\Theta_{th}}\geq 1, 
\label{eq:break}
\end{equation}
where $\varepsilon$ denotes the longitudinal strain, 
$\Theta_i$ and $\Theta_j$ are the generalized bending angles at the
two beam ends, and $\varepsilon_{th}$ and $\Theta_{th}$ are breaking
thresholds which have fixed values for all the beams. The two terms of
Eq.\ (\ref{eq:break}) characterize the 
contributions of the 
stretching and bending failure modes of a beam. 
Simulations showed that the local shear of
the particle contacts provides the main contribution to the bending
angles $\Theta_i$, $\Theta_j$, so that bending dominated beam breaking
in Eq.\ (\ref{eq:break}) characterizes crack formation due to
shear. 
Varying the values of the breaking thresholds $\varepsilon_{th}$ and
$\Theta_{th}$, the relative importance of stretching and bending can
be controlled: increasing the value of a breaking parameter, the
effect of the corresponding failure mode diminishes. 
An important feature of the breaking criterion is that in Eq.\
(\ref{eq:break}) the deformation $\varepsilon$ is not 
restricted to positive values contrary to former simulation studies on
brittle fracture and fragmentation
\cite{wittel_prl_2004,wittel_shape_prl,astrom_branching_1997}. Since
the first term of  
Eq. (\ref{eq:break}) becomes negative 
when the beam is compressed, failure is dominated by the bending/shear
mode in such a way that increasing compression increases the shear
resistance of the beam. 
\begin{figure}
\begin{center}
\epsfig{bbllx=25,bblly=435,bburx=690,bbury=720,file=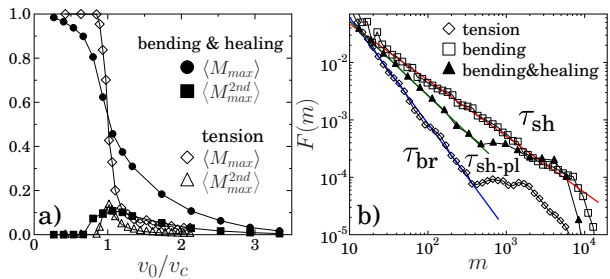,
  width=8.2cm}
 \caption{\small
{\it (Color online)} $(a)$ The average mass of the largest
$\left<M_{max}\right>$ and second largest $\left<M^{2nd}_{max}\right>$
fragments as a function of the impact velocity $v_0$ normalized by the
critical velocity $v_c$ of fragmentation for two parameter sets. $(b)$
Fragment mass distributions obtained for the limiting 
cases of tension, bending, and bending\&healing dominated
breakups giving the corresponding exponents $\tau_{br}$,
$\tau_{sh}$, and $\tau_{sh-pl}$, respectively.
}
\label{fig:largest}
\end{center}
\vspace*{-0.5cm}
\end{figure}
In order to represent the plastic behavior of the material, 
we assume that the beams have a linear elastic behavior 
up to fracture, but, whenever two particles are pressed
against one another for a time longer than $t_h$, a new, undeformed beam
is inserted between  
them. This way during the impact process, the
particle contacts may undergo a sequence of breaking-healing events
which leads to plastic energy dissipation 
and to the appearance of permanent deformation.
Varying the healing time $t_h$ the mechanical response of the
model material can be controlled: $t_h = 0$ corresponds to the case of
perfect shear plasticity, while $t_h \to \infty$ implies no
healing at all, i.e.\ brittle behavior.

In order to investigate the effect of local failure modes of beams on
the fragmentation process,  we carried out computer simulations by
setting the stretching threshold to a fixed value $\varepsilon_{th} =
0.02$ and varying the bending threshold within a broad range  $1.0\leq
\Theta_{th} \leq 200$. In this 
way $\Theta_{th} = 200$ and $\Theta_{th}=1$ imply total tension
and bending dominance, respectively, while intermediate $\Theta_{th}$
values interpolate between the two limits. 
\begin{figure}
\begin{center}
\epsfig{bbllx=20,bblly=20,bburx=550,bbury=395,file=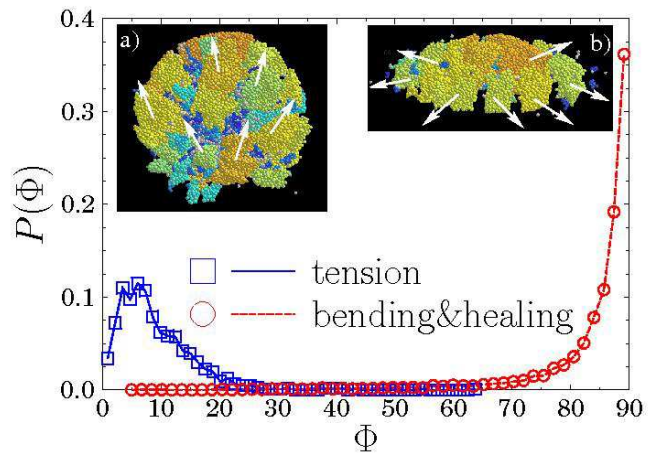,
  width=8.5cm}
 \caption{\small
{\it (Color online)} Probability distribution of the angle $\Phi$
between the velocity vector of fragments and the direction of impact
for tension $\Theta_{th}=200$ and shear $\Theta_{th}=1.0$ dominated
breakup. Insets: final states of fragmentation for tension $(a)$ and
shear $(b)$ dominance. The arrows indicate the direction of the
velocity of a few fragments. }
\label{fig:final}
\end{center}
\vspace*{-0.5cm}
\end{figure}
Simulations were stopped when the system
reached a relaxed state, i.e.\ no beam breaking occurred during 1000
consecutive time steps. 
Figure \ref{fig:exp_deform}$(b)$ presents the final state of an impact
simulation obtained with a sample of $N=24000$ particles at 
low impact velocity. It can be observed that the model is able to
reproduce both the deformation and the crack structure of PP,
with parameter values where the beam breaking is dominated by
bending $\Theta_{th}=1.0$, 
furthermore, compressed contacts 
easily heal $t_h=0$. The large permanent deformation of the sphere in
Fig.\ \ref{fig:exp_deform}$(b)$ arises due to breaking-healing
sequences of particle contacts in the compressed impact zone.
Above this zone tensile stresses arise resulting in opening
cracks along the impact direction in agreement with the experiments.  
Figure \ref{fig:largest}$(a)$
presents the average mass of the largest $\left<M_{max}\right>$ 
and second largest  $\left<M^{2nd}_{max}\right>$ fragments as a
function of the impact velocity $v_0$ for the
tension $\Theta_{th}=200$ dominated brittle
breakup $t_h=\infty$ ({\it tension}), and
for the bending dominated fragmentation $\Theta_{th}=1.0$
with perfect shear plasticity $t_h=0$ ({\it bending\&healing}). 
For low impact velocities the sample gets only damaged, hence, the
largest fragment comprises nearly the entire mass $M_0$ of the body
$\left<M_{max}\right> \sim M_0$, while the second largest one is
orders of magnitude smaller  $\left<M^{2nd}_{max}\right> \ll
\left<M_{max}\right>$. Fragmentation occurs when the largest and
second largest pieces become comparable, i.e.\ at the critical
velocity $v_c$ where the $\left<M^{2nd}_{max}\right>$ curve has a
maximum coinciding with the inflexion point of
$\left<M_{max}\right>$. Figure \ref{fig:largest}$(a)$ shows that for
the breakup of brittle materials dominated by tensile stresses, the
damage-fragmentation transition is sharp in agreement with
experiments \cite{katsuragi_prl_2005,glassplate_kadono_1997}, however,
when shear breaking dominates a 
broad critical regime emerges.
The reason is that at $v_c$ the largest fragment does not break up into
large pieces as for the tension case, but instead gradually erodes with a
cleavage mechanism giving rise to a slowly decaying residue.

Representative examples for the mass distribution of fragments are
presented in Fig.\ \ref{fig:largest}$(b)$ for three
parameter sets. For the mass distribution of the tension dominated
breakup of heterogeneous brittle
materials obtained at the 
parameter values $\Theta_{th}=200$, $t_h=\infty$ ({\it tension}) a
power law behavior is evidenced with the usual exponent
$\tau_{br}=1.9\pm 0.1$ \cite{kun_fragment_pre2008}.  
Simulations showed that decreasing the healing time $t_h\to 0$ in the
tension limit practically does not affect the
fragmentation process because fragments are only generated by opening
cracks which do not let healing play any role. It is important to
emphasize that the breakup process 
substantially changes when shear dominates the crack formation
$\Theta_{th}=1$  ({\it bending}). Even in the case of perfectly
brittle beam breaking 
$t_h=\infty$ the low shear resistance leads to a
significantly lower 
exponent than in the tensile limit $\tau_{sh}=1.0\pm 0.05$ (see Fig.\
\ref{fig:largest}$(b)$).  Increasing the strength
of plasticity  $t_h \to 0$ when shear dominates the cracking
$\Theta_{th}=1$, the exponent of the
fragment mass distribution slightly increases: due to the healing of cracks
fragments can merge which decreases
the relative frequency of large pieces leading to a faster decay of
the distribution. The highest value of
the exponent $\tau_{sh-pl}=1.25\pm0.06$ is obtained in the plastic limit
$t_h=0$ together with $\Theta_{th}=1.0$ ({\it bending\& healing}). 
It is important to emphasize that varying solely the velocity of
impact in Fig.\ \ref{fig:exp_mass}, 
for the mass distribution of fragments an excellent agreement is
obtained between the shear-plastic simulations and the
experiments. 
The results show that the shear dominated
cracking together with the healing mechanism of compressed crack
surfaces are responsible for the unique fragmentation of plastic
materials. 
During the impact process of the experiments a considerable fraction
of the kinetic energy of the particle gets transformed into heat
which enhances the effect of healing and plasticity in consistence
with our theoretical results.
Our extensive simulations indicate that the exponent
$\tau_{sh-pl}$ is universal, i.e.\ it does not depend on the impact
velocity or on materials' microstructure
characterizing a novel universality class of 
fragmentation phenomena.    

Due to low shear resistance, we find
that the fragmentation of plastic shows similarities to the breakup of
liquid droplets colliding with a wall.
The spatial distribution of fragments and
the crack structure in the final state of brittle and plastic
fragmentation are compared in the insets of Fig.\
\ref{fig:final}. Although meridional and segmentation cracks are
clearly observed in the brittle case in agreement with experiments, 
a lateral spreading of fragments is obtained for plastic with shear
dominated breaking. For the quantitative characterization of the
spatial spread of fragments,  the distribution
of the angle $\Phi$ of the velocity vector of fragments with the
impact direction (vertical) is presented in Fig.\ \ref{fig:final}. It
can be observed that most 
of the fragments bounce back from the hard wall when tension
dominates, however, in the shear-plastic case the fragments escape
laterally producing a ``splash'' of the entire body similar to
liquid droplets \cite{villermaux_2007,PhysRevE_moukarzel}. 

In conclusion, our experimental and theoretical study revealed that
the breakup of plastic materials falls into a novel universality class of
fragmentation phenomena characterized by the new mass
distribution exponent. Based on discrete element simulations we showed
that the plastic behavior of the material together with the dominance
of shear in crack formation are responsible for the
substantial difference from brittle
fragmentation. The low shear resistance of the material gives rise to
a splashing similar to the breakup of droplets of highly
viscous liquids. Beyond the industrial importance of the fragmentation
of polymeric materials, our results might also be applied to obtain a deeper
understanding of the fragmentation of highly viscous magma during
pyroclastic activity at volcanic eruption
\cite{fragment_magma_nature373,fragment_magma_nature380}. 
For theoretical investigations our results
demonstrate that the breakup of 
solids cannot be understood as a generic stochastic process since the
precise mechanism of crack initiation and growth, i.e.\ the dominance of
tensile or shear stresses govern the exponent of
fragmentation. Still remains the challenge to construct a 
theoretical approach which explains the emergence of universality
based on crack dynamics.

\begin{acknowledgments}
The project was partially supported by the German Federal Ministry 
of Economics and Technology  (BMWi) via AiF-Grant-No. 14516N.  
F.\ K.\ acknowledges the financial support of the Janos Bolyai
Grant of HAS. 
\end{acknowledgments}

\bibliography{/home/feri/MAGAN/MAGYAR/MTA_DOKTORI/statphys_fracture}

\end{document}